# On the mechanism of hydrophilicity of graphene

Guo Hong[1], Yang Han[2], Thomas M. Schutzius[1], Yuming Wang[3], Ying Pan[1], Ming Hu[2, 4], Jiansheng Jie[3], Chander S. Sharma[1], Ulrich Müller[5], Dimos Poulikakos[1]*

[1]Laboratory of Thermodynamics in Emerging Technologies, Department of Mechanical and Process Engineering, ETH Zurich, Sonneggstrasse 3, 8092 Zurich, Switzerland

[2]Institute of Mineral Engineering, Division of Materials Science and Engineering, Faculty of Georesources and Materials Engineering, RWTH Aachen University, 52064 Aachen, Germany

[3]Institute of Functional Nano & Soft Materials (FUNSOM), Collaborative Innovation Center of Suzhou Nano Science and Technology, Jiangsu Key Laboratory for Carbon-Based Functional Materials & Devices, Soochow University, Suzhou, Jiangsu 215123, P. R. China

[4]Aachen Institute for Advanced Study in Computational Engineering Science (AICES), RWTH Aachen University, 52062 Aachen, Germany

[5]Empa, Swiss Federal Laboratories for Materials Science and Technology, Laboratory for Nanoscale Materials Science, Überlandstrasse 129, CH-8600 Dübendorf, Switzerland

ABSTRACT

It is generally accepted that the hydrophilic property of graphene can be affected by the underlying substrate. However, the role of intrinsic vs. substrate contributions and the related mechanisms are vividly debated. Here we show that the intrinsic hydrophilicity of graphene can be intimately connected to the position of its Fermi level, which affects the interaction between graphene and water molecules. The underlying substrate, or dopants, can tune hydrophilicity by modulating the Fermi level of graphene. By shifting the Fermi level of graphene away from its Dirac point, via either chemical or electrical voltage doping, we show enhanced hydrophilicity with experiments and first principle simulations. Increased vapor condensation on graphene, induced by a simple shifting of its Fermi level, exemplifies applications in the area of interfacial transport phenomena.

Graphene has attracted great research attention due to its extraordinary electrical, mechanical and optical properties[1-3]. However, its wetting behavior with respect to water, which can be relevant to many applications,[4-7] is still not well understood. For several years, graphene has been regarded as a hydrophobic surface, and its wetting properties were considered similar to those of graphite[8,9]. Recent studies suggest that graphene on copper is initially more hydrophilic, and airborne contaminants[10,11] or water adsorption[12] are responsible for its perceived hydrophobicity. However, even with exercising care to generate clean (no hydrocarbon adsorption) graphene surfaces, different values of static water contact angles (WCA) have been reported on variety of substrates, like copper (44º) and highly ordered pyrolytic graphite (HOPG) (60º)[13], which implies that the underlying substrate does play a role in the hydrophilicity of graphene. Current research focuses on determining whether water molecules on graphene can interact directly with the underlying substrate through such an atomically thin material, leading to a degree of graphene "transparency" to wetting. Two major conclusions were drawn recently, ranging from transparency[14] (graphene is invisible to the wetting behavior of the underlying substrate), to transluceny[15] (graphene is more transparent to wetting on hydrophilic substrates and more opaque to wetting on hydrophobic substrates). Such diverging conclusions, which may also be dependent on the experimental conditions investigated, clearly indicate that our understanding of the physics of graphene wetting needs to be further advanced, to enable its control and, through this, the engineering of desirable wetting performance of this important novel material.

In order to perform respective wetting angle measurements to study the intrinsic wettability of graphene, well-controlled modifications (different substrates or chemical doping) are highly desired. Unfortunately, on one hand, high quality single layer graphene can only be grown on specific kinds of substrates according to the state of the art[16,17]. Although, in principle, graphene can be transferred to any substrate afterwards, unintended contamination (e.g. polymer residue) is experimentally unavoidable, which seriously hinders investigations. On the other hand, chemical doping has been comprehensively developed in the past years[18-20], but the exposed graphene surface is normally occupied by the introduced heteroatoms, affecting the wetting property.

To study the effect of chemical doping on the interaction between graphene and water, we modified graphene according to the following procedure. First, we introduced graphene—on a supporting Cu substrate—into a fluorination chamber, which cleaned and chemically functionalized a single side of the graphene sheet; here, the supporting substrate acted as a protective mask for the other side. Then, we protected this chemical functionalization on graphene with a polymer layer, deposited by spin-coating, and removed the underlying growth substrate, revealing a graphene surface again (we term this "created" surface)[6]. This process enabled us to place the chemical modifications of graphene at the buried interface between the polymer layer and graphene and not on the exposed surface. Next, we made sure through measurements (to be detailed later) that such exposed surfaces ("created" surfaces) were saturated with hydrocarbons before evaluation with respect to wetting. This allowed us to achieve stable samples (removing the effect of time-dependent adsorption) and reproducible data during repeated measurements. In addition, it is expected that in applications graphene will indeed be exposed to hydrocarbons in the atmosphere and reach naturally a saturated state.

Following this approach to study wetting, single layer graphene films (Fig. S1) were synthesized by chemical vapor deposition (CVD) on copper foils. In order to chemically functionalize graphene, gas phase fluorination was performed (Xactix® $XeF_2$ etching system), producing a sparse, atomically thick layer of fluorine.[21] Each exposure cycle of the $XeF_2$ gas is set at 30 s, and the total fluorination time was controlled for 8, 24 and 300 cycles at 30 °C. After that, a thin layer of poly(methylmethacrylate) (PMMA) was spin-coated on top, to protect the fluorination layer. The thickness and uniformity of PMMA is controlled by the rotational speed of the spin coater and the concentration of the PMMA/solvent solution. The copper substrate was subsequently etched away in a $(NH_4)_2S_2O_8$ solution (0.5 M in water)[22]. The concentration of the etchant in solution is ensured to be the same for each sample. The entire etching process took ~ 2 hours, and then the samples were dried by blowing with nitrogen gas. The WCA measurements were subsequently performed on this "created" graphene surface. XPS spectroscopy showed that the "created" graphene surface was saturated with hydrocarbons, and the spectra of "created" and "aged" graphene surfaces overlap (Fig. S2). "Aged" graphene surfaces were kept in open air for ~1 day, which in previous studies has been shown to be sufficient for the surface to be completely saturated with hydrocarbons[10].

Figure 1 compares the advancing, static and receding WCAs of graphene layers for different substrate configurations. For the advancing and receding WCA measurements, videos of adding or withdrawing deionized water to and from an existing water droplet were recorded by a CMOS camera attached to an objective. The advancing (receding) WCA was determined by measuring the values when the contact line of water droplet moved while adding (withdrawing) water. For the aged graphene on copper foil (No. 1 in Fig. 1), these values were found to be 84°, 80° and 46°, respectively (the wetting behavior of clean graphene on copper foil can be found in Supplementary Fig. S6). After 4 min of fluorination, graphene supported by a copper substrate (No. 2 in Fig. 1) had a significant increase in its advancing and static WCAs, because the fluorine terminal has been confirmed to be hydrophobic[23]. Following the above approach, the created surfaces of the non-functionalized (No. 3 in Fig. 1) and backside-fluorinated graphene (No. 4 in Fig. 1) were measured, and the latter was found to be more hydrophilic. Since PMMA was used as a substrate for all samples, the changes of the WCAs can only be attributed to the backside fluorination. By extending the fluorination time to 12 min (No. 5 in Fig. 1), the surface hydrophilicity was further enhanced. The case of aggressive fluorination (150 min; No. 6 in Fig. 1) will be discussed later. Fluorine is acting directly on graphene and is not a passive substrate layer. To explain the observed behavior, we hypothesized that the chemical doping (fluorination) can shift the Fermi level of graphene, which could have further implications on wetting. The Fermi level shift of graphene by fluorination is confirmed by DFT simulation in Supplementary Fig. S3 and S4.

To further underpin the validity of this hypothesis, a DC voltage bias was applied to graphene samples on a $SiO_2$/Si substrate (transferred from copper foil), since electrical voltage doping has the equivalent effect as chemical doping but is a cleaner and more controllable process. Moreover, the Fermi level shifting can be easily determined when subjecting the sample to the applied electric field.[24-26] After transfer, all the graphene samples were annealed in a 10% hydrogen/argon atmosphere at 500 °C for one hour to remove any possible polymer residue. They were then stored at ambient conditions for one week to reach saturation state with respect to carbon species (minimization of the adsorption process during later measurements). Considering the high voltage applied on the sample, only static WCA measurements were performed for safety reasons. The measurement setup is described in Fig. 2a, in which graphene on a $SiO_2/Si$ substrate was attached to a cooper plate by silver paste. The copper plate served as the bottom electrode and a glass slide was used to isolate it from the other components of the setup. The top electrode was a copper wire fixed on the graphene layer by silver paste. A DC voltage source was used to apply the high voltage across the sample. A current meter was used to make sure that there was no current shorting through the dielectric layer. Specifically, the droplet was carefully deposited on the graphene surface at least 3 mm away from the top electrode to avoid electrical wetting interference[27]. The static WCA of "aged" graphene samples was found to be 88°, which confirms that the hydrocarbon adsorption was saturated. When a negative voltage of 100 V was applied (the thickness of $SiO_2$ layer is 285 nm, which one can use to estimate an electrical potential of $3.5*10^8$ V/m), which is equivalent to n-doping of graphene, the Fermi level moves upwards with respect to the Dirac point and the WCA decreased to 78°, as expected by the π-H model[28,29]. Interestingly, when a positive voltage of 100 V was applied, which leads to the p-doping of graphene, the WCA decreased even more, to 60° (Supplementary Movie 1). Both negative and a positive voltage biases led to a stronger interaction between water and graphene and enhanced wetting, the reason for which will be explained below with the help of theoretical calculations. Also, for both cases, the WCA values stayed constant after turning off the bias. This is because both values were larger than the receding angles leading to hysteresis. Reference measurements were performed on copper/$SiO_2$/Si (Fig. 2b) and $SiO_2$/Si (Fig. 2c) samples as well, under the same parameters. It was found that the WCAs stayed the same under different applied voltage biases (-100 V/0 V/+100 V) in both cases. These results support the hypothesis that the intrinsic hydrophilicity of graphene is related to its Fermi level state and not to charge trapping at the interface between the adsorbed hydrocarbons on graphene and water.

In order to investigate the physics behind the experimental observations more directly, DFT simulations were performed to study the interaction between a water molecule and graphene. Two preferable configurations, one- and two-leg models, respectively, were considered in detail (Supplementary Section SI-1 to SI-3). Each configuration was fully optimized under different conditions, *i.e.*, neutral, -1e and +1e (graphene cell doped by one hole or electron, respectively), as illustrated in Fig. 3. To quantitatively describe the interaction between a water molecule and graphene, their distance (d) and the corresponding adsorption energy (ΔE) were calculated and are given in Table 1. For the neutral one-leg configuration, these results are d = 3.287Å and ΔE = -0.136 eV. The dipole $\vec{P}$ of the water molecule oriented itself at an angle θ ≅ 127° with respect to *z*-axis (left panel in Fig. 3a). However, when the graphene was charged by -1e or +1e, the corresponding values of d and ΔE adjusted themselves accordingly. To this end, when one electron was subtracted (-1e), the dipole of the water molecule oriented itself upwards (θ = 0°), i.e. perpendicular to the graphene plane (middle panel in Fig. 3a). Both d = 3.066 Å and ΔE = -0.305 eV became smaller compared to the values of the neutral state, indicating that the interaction between water molecule and graphene had been strengthened. In contrast, when one electron was added (+1e), the dipole $\vec{P}$ became antiparallel with respect to *z* axis (θ = 180°, right panel in Fig. 3a). Also in this case, d = 3.173 Å and ΔE = -0.303 eV were smaller compared to the neutral state above (stronger interaction). Similar phenomena were also found for the two-leg configuration: the water molecule was shown to flip to its preferable orientation under electron- or hole-rich conditions. Both upward and downward rotations led to smaller d and ΔE and more energetically favorable configurations. As a result, the doped graphene-water interaction is a more stable configuration, i.e., lower energy state (Fig. 3 and Table 1). Thermodynamically, this equates to a stronger interaction between water and graphene, which would lead to a reduced contact angle.

We further analyzed why both the one- and two-leg water molecule models flipped to the same orientation for the same charge state of graphene. Positively charged graphene attracts the oxygen atom of the water molecule (middle panel in Figs. 3a, 3b). Similarly, negatively charged graphene attracts the hydrogen atoms of the water molecule (right panel in Figs. 3a, 3b). The summation of Coulomb interactions and van der Waals forces decides the final configuration (Supplementary Fig. S5). As a result, when graphene is supported by a substrate or it is doped (either chemically or through an electrical voltage), its electron/hole density of states increases accordingly and leads to a more hydrophilic surface. On the other hand in the case of saturated fluorination of 25% coverage, the doping can break the band structure of graphene turning it from semi-metal to insulator[21]. This leads to a more hydrophobic surface (No. 6 in Fig. 1). Although the intrinsic wetting of clean suspended graphene is not directly measurable due to inherent experimental difficulties, its behavior is expected to be more hydrophobic than that of supported graphene on a copper foil reported previously[10], since the graphene layer is initially n-doped by the copper substrate[30]. To support this statement, we also prepared graphene samples on copper foil in CVD and cooled them down in argon or hydrogen atmospheres, respectively. It was found that graphene cooled down in an argon atmosphere led to higher advancing and receding WCAs compared to those of graphene cooled down in a hydrogen atmosphere (Fig. S6). This is because graphene was n-doped when deposited on a copper substrate, and the Fermi level will be shifted upwards with respect to the Dirac point, which would render the surface more hydrophilic. If a minute amount of $O_2$ exists in the growth process, it will have a p-doping effect, potentially counteracting the n-doping effect of the copper substrate[31]. (Note that unlike the argon atmosphere, the hydrogen atmosphere consumes the oxygen traces at high temperatures.) Recently, it was reported that double layer graphene is more hydrophobic than single layer graphene[32]. In this configuration, the doping effect of the substrate will be weakened by additional graphene layers[33,34]. As a result, a single layer graphene is expected to experience a stronger doping effect from the substrate resulting in a larger Fermi level shift from its Dirac point, leading to enhanced hydrophilicity.

Through this work, we showed that by tuning the Fermi level of graphene—through either chemical or electrical doping—we can tune the wetting behavior of water to it. We now aim to show through an illustrative example application where this tunable wetting behavior can be beneficial (condensation), and how such findings can affect important interfacial transport phenomena with respect to utilizing graphene. Recently, graphene has been shown to be a promising material for promoting efficient dropwise condensation,[35] an important energy application,[35-40] and understanding and controlling its wetting property with respect to water is extremely important for the overall process efficiency. Figure 4a shows a schematic of an experimental setup we employed to study water vapor condensation on graphene. A standard cooling stage was used for the water condensation on graphene in open air. Thermocouples were used to record the temperature of the stage and of the graphene surface, respectively. Vaporizing liquid nitrogen was enclosed and circulated under the sample stage. The configuration of the graphene sample is exactly the same as in the water contact angle measurements with the bias as described earlier (Fig. 2). An optical microscope was used to observe the condensation phenomenon. Corresponding videos were recorded by a CMOS camera. The first frame before water started to condense was set as the condensation starting point (0 s). When the surface is cold enough (5 $^{o}$C with 24% relative humidity in open atmosphere), water droplets start to condense. Figures 4b-c show the condensation phenomenon at the same position of a graphene sample with and without an applied voltage bias. The resulting condensate droplet numbers and radius are plotted on the left hand side of the frames of each reported condensation time. After 25 s, for example, it is found that there were visibly more active nucleation sites on the more hydrophilic graphene sample (with electrical voltage doping of 80 V). By prolonging the cooling time, droplets grew gradually and small droplets merged into larger ones through Oswald ripening (50 s of Figs. 4b and 4c). It is well known that an important parameter in controlling the nucleation density of water on a surface is its intrinsic contact angle (barrier to nucleation). At all times, there were more water droplets observed on the more hydrophilic graphene with applied bias, demonstrating enhanced condensation (Supplementary Movie 2 shoes the early stages of condensation, where the differences are visible).

A condensation process was also investigated in the controlled environment of an environmental scanning electron microscope (ESEM) for the backside-fluorinated graphene samples described previously (No. 5 in Fig. 1). The samples were bonded to an aluminum strip using a silver paste for the water condensation experiments in the ESEM (FEI Quanta 600). The aluminum strip was attached to the cooling stage of the ESEM, which was fixed to be 2 $^{o}$C with a maximum temperature fluctuation of 0.1 $^{o}$C (reported from a temperature sensor). The ESEM chamber was evacuated first and then the system pressure was increased slowly by adding water vapor until the water started to condense on the graphene surface. Once the first droplets appeared on the graphene surface, the system pressure was kept constant and the droplets grew slowly (Supplementary Movie 3 and Fig. S7). The condensation process was scanned and recorded every 2.84 seconds for each frame. Figure 4d is a typical ESEM image of a "created" graphene surface before condensation starts (tilted view of 86.5$^{o}$). It was found that condensation required markedly higher (0.2 torr) pressure to initiate on the non-functionalized, less hydrophilic graphene (Fig. 4e) compared to the backside-fluorinated graphene (Fig. 4f) Additionally, the advancing WCA of the growing droplets was smaller for the backside fluorinated graphene (insert in Fig. 4e and 4f), all of which further underpins the claim that a Fermi level shift is responsible for increased graphene hydrophilicity (Supplementary Movie 3).

We have shown that an underlying substrate or dopants can indeed affect the hydrophilicity of graphene, but the reason does not appear to be the van der Waals interaction between water molecules and substrate through the effectively "transparent" graphene layer as claimed previously[14,15] because in all our experiments, the substrates were the same (PMMA in chemical doping and SiO$_2$ in electrical doping). The underlying substrate or dopants (either p- or n-doping) modify instead the Fermi level of graphene and therefore its interaction with the polar water molecules, which respond to such Fermi level changes. DFT simulations indicate that the configuration of a water molecule will adjust itself to the doping states of graphene. Accordingly, the hydrophilicity of graphene can be tuned with such doping. Our results provide new evidence in terms of explaining the underlying physics and quantifying the hydrophilicity of graphene. As an example application, enhanced water condensation was demonstrated on graphene with increased hydrophilicity, obtained simply by an applied electrical voltage bias or chemical doping.

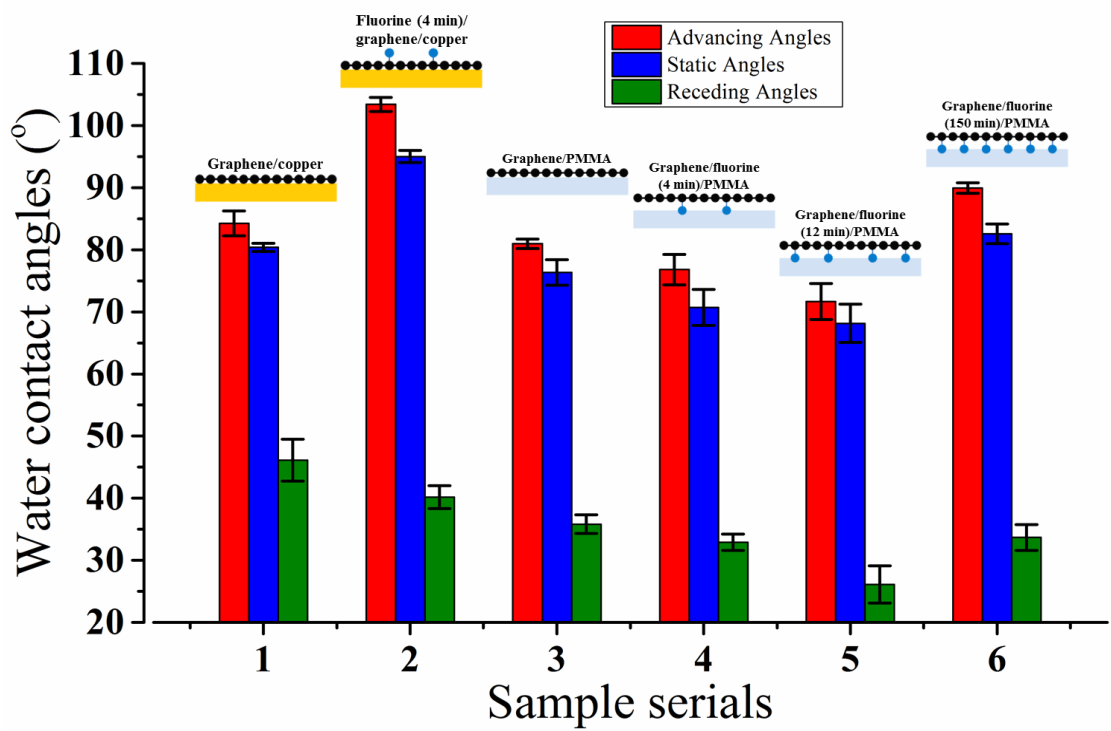

**Figure 1.** Evolution of the advancing/static/receding water contact angles of graphene during single side fluorination. (1) Aged graphene on copper foil. (2) 4 min fluorination of graphene on copper foil. (3) Graphene on PMMA. (4) Graphene with 4 min fluorination of its backside on PMMA. (5) Graphene with 12 min fluorination of its backside on PMMA. (6) Graphene with 150 min fluorination of its backside on PMMA. The measured graphene layers in samples 2-6 were created according to the processes explained in detail in the text. Each sample was measured 5 times and reported are the average values. The error bars were the standard deviation of the results. Droplets in contact with real surfaces in an air environment do not form a unique contact angle with the surface. Rather, they can have a spectrum of values that lie between an upper (advancing contact angle) and lower (receding contact angle) bound. The values that lie in the range between those two bounds are called static contact angles. The advancing and receding angle together show the mobility of droplets on a surface and the receding angle alone shows the adhesion of the droplet to the surface. For completeness, we also report the static contact angle.

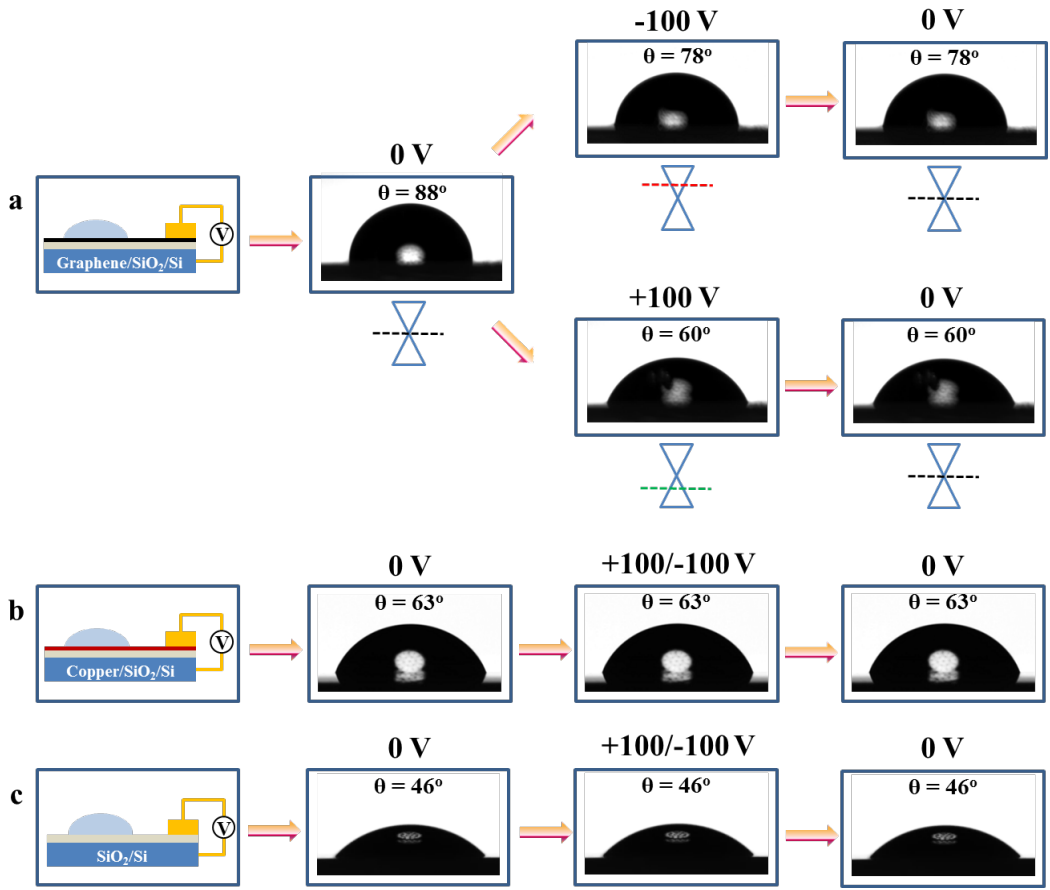

**Figure 2.** Wettability of different graphene surfaces with bias. (**a**) Wettability of graphene transferred on $SiO_2$/Si substrate. The silicon substrate served as the bottom electrode while the top electrode was a copper wire fixed by silver paste on the graphene. The droplet was deposited on the graphene surface at least 3 mm away from the top electrode. At neutral state, the static water contact angle was close to 88°. In the presence of both negative and positive bias, the water contact angle decreased. After turning off the bias, the water contact angle remained the same for both cases. (**b**) Wettability of copper on $SiO_2$/Si substrate. (**c**) Wettability of $SiO_2$/Si surface. At different applied voltages (-100 V/0 V/+100 V), the water contact angles remained unchanged.

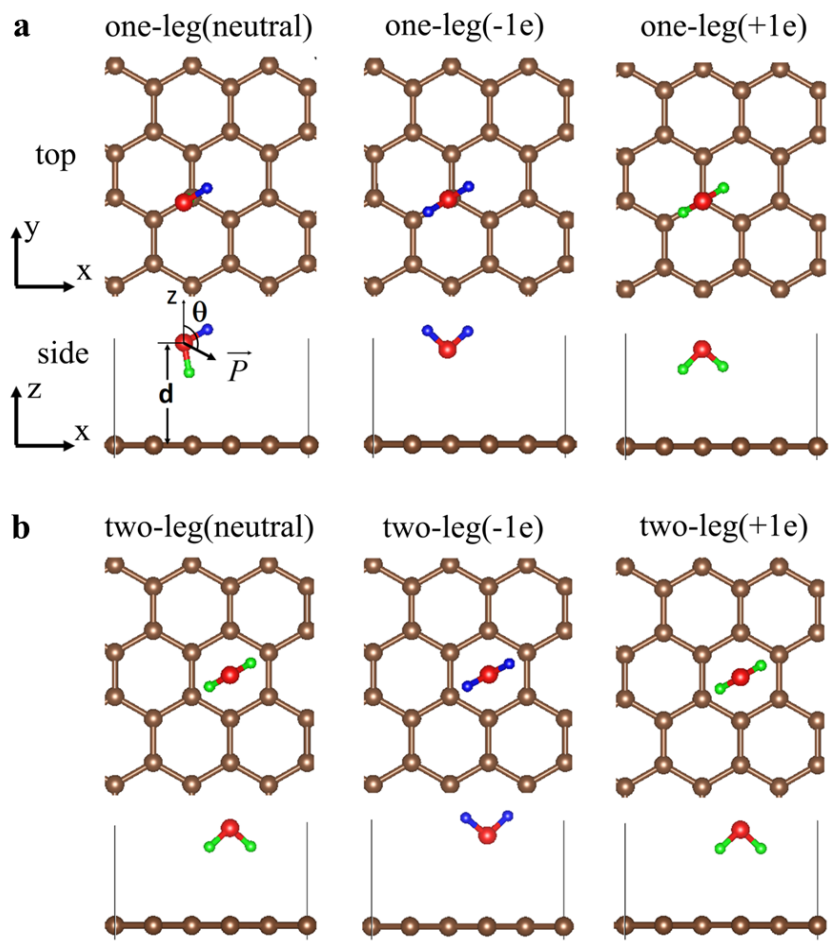

**Figure 3.** Optimized configurations of a water molecule on suspended graphene. (a, b**)** One (top raw)- and two-leg (bottom raw) models of top and side views of a water molecule on graphene in neutral (left), -1e (middle) and +1e (right) states, respectively. The large brown and red balls represent carbon and oxygen atoms, respectively. To show clearly the different distances of the two hydrogen atoms in the water molecule from the graphene, the small blue and green balls denote the hydrogen atoms away and toward the graphene layer, respectively. In the left panel of (a), d, θ and $\vec{P}$ have been labeled, representing the distance between water molecule and graphene, the angle of the dipole of the water molecule with respect to z axis and the dipole of the water molecule.

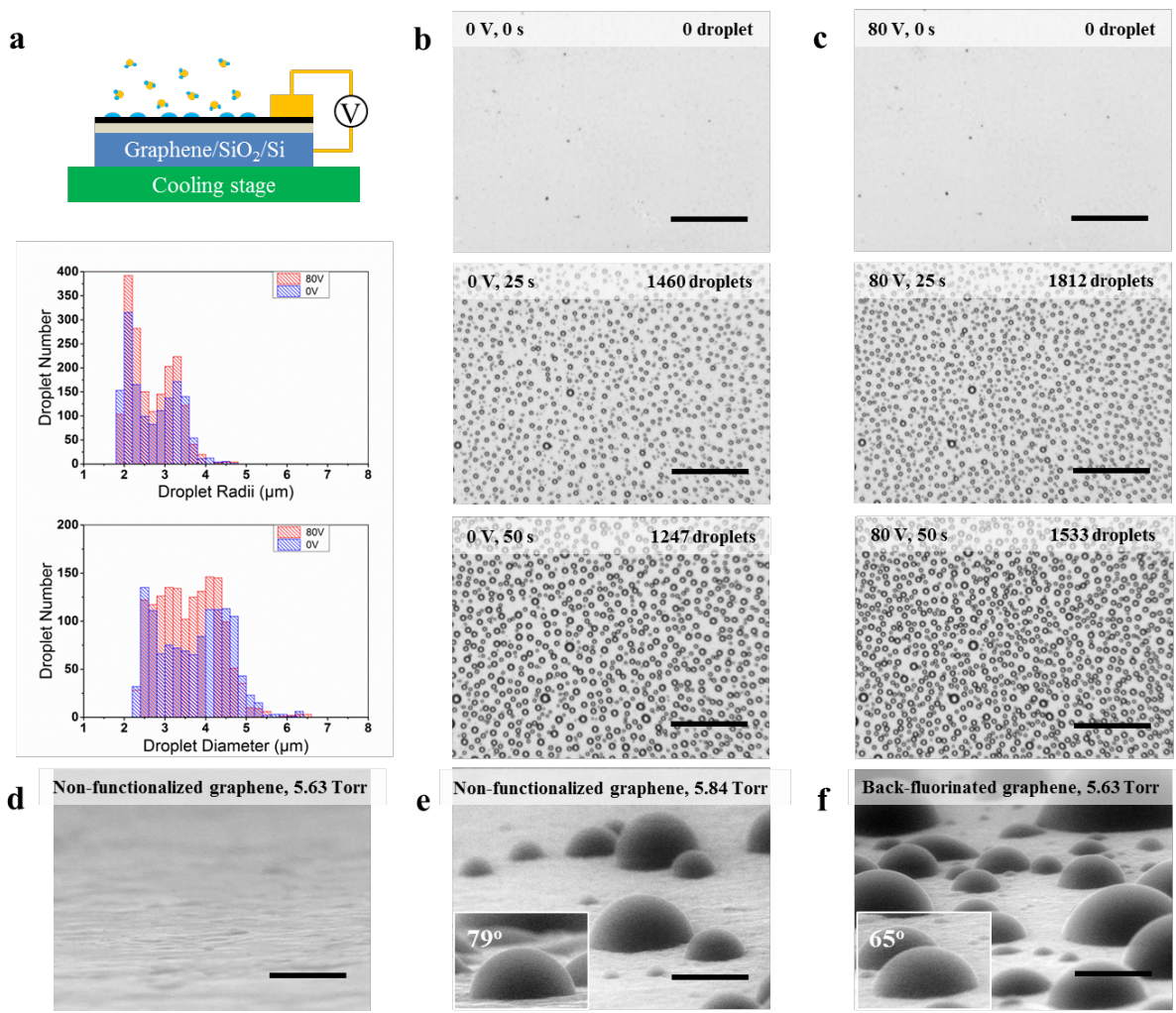

**Figure 4.** Water condensation on graphene. **(a)** Schematic of the setup of the water condensation experiment. (b, c**)** Temporal evolution of water droplets condensed on the same position of a graphene sample with and without voltage doping, respectively. For the case of 80 V positive voltage doping, more water droplets were found to condense on graphene. Videos were recorded for the condensation process. One frame before water condensation initiation was set as the start point (0 s). Corresponding droplet numbers and radii are shown on the left hand side of each condensation time. (d**)** Typical ESEM image of non-functionalized graphene before the initiation of condensation. (e, f**),** ESEM images of water condensed on non-functionalized and backside fluorinated graphene, respectively. The inset images contain the advancing WCA of the growing droplets. The scale bars are 100 μm for (b) and (c), 20 μm for (d), (e) and (f).

**Table 1.** Parameters of optimized configurations of a water molecule on graphene. For both one- and two-leg models, the distances and adsorption energies between water molecule and the charged graphene at charged states (-1e and +1e) decrease compared to the neutral state, which indicates a stronger interaction between water and graphene.

| Models | One-leg | | | Two-leg | | |
|---|---|---|---|---|---|---|
| Configurations | neutral | -1e | +1e | neutral | -1e | +1e |
| Distance (Å) | 3.287 | 3.066 | 3.173 | 3.152 | 2.943 | 3.099 |
| ΔE (eV) | -0.136 | -0.305 | -0.303 | -0.146 | -0.314 | -0.323 |

## ASSOCIATED CONTENT

Supporting Information

The Supporting Information is available free of charge on the ACS Publications website at DOI:

XPS spectroscopy, First principle simulation discussion, temporal evolution of the advancing and receding water contact angles measured on graphene on copper foil, pressure variation in the ESEM, real-time movie of static water contact angle of graphene on $SiO_2$/Si substrate with +100 V bias, real-time movie of water droplets condensed on graphene/$SiO_2$/Si samples in the open air with and without applied electrical voltage bias of 80 V, movie of water droplets condensed on graphene (non-fluorinated; PMMA substrate) and backside-fluorinated graphene (PMMA substrate).

## AUTHOR INFORMATION

**Corresponding Author**

*E-mail: dpoulikakos@ethz.ch

**Notes**

The authors declare no competing financial interest.

## ACKNOWLEDGEMENT

Financial support for this project is provided by ETH Zurich, Swiss National Science Foundation (200021_146898/1, 200021_146180/1 and 200021_144397/1) and European Research Council Advanced Grant (669908 INTICE). T.M.S. gratefully acknowledges the ETH Zurich Postdoctoral Fellowship Program and Marie Curie Actions for People COFUND program (FEL-14 13-1). We thank Ms. Asel Maria Aguilar Sanchez and Ms. Gabriele Peschke from the Institute for Building Materials, ETH Zurich for the support of the ESEM measurements.